\documentclass{nature}

\newcommand\ulap[1]{\vbox\@to\z@{{\vss#1}}}% 
\newcommand\dlap[1]{\vbox\@to\z@{{#1\vss}}}% 

\usepackage{graphicx}
\usepackage{tabularx}
\usepackage{multirow}
\usepackage[normalem]{ulem}
\usepackage{textcomp,gensymb}
\usepackage[version=4]{mhchem}

\newcommand{\ci}{[C\,{\sc i}]\phantom{.}(1--0)}
\newcommand{\cii}{[C\,{\sc i}]\phantom{.}(2--1)}

\newcommand*\farcs{\ensuremath{\overset{\prime\prime}{.}}}

\usepackage[colorlinks = true,
            linkcolor = black,
            urlcolor  = blue,
            citecolor = black,
            anchorcolor = blue]{hyperref}
            
\usepackage[dvipsnames]{xcolor}

\def\farcm@apj{% 
 \mbox{.\kern -0.7ex\raisebox{.9ex}{\scriptsize$\prime$}}% 
}
\usepackage{lineno}

\title{The ramp-up of interstellar medium enrichment at z\,$>$\,4}

\author{M.~Franco$^{*1}$, K.\,E.\,K.~Coppin$^1$, J.\,E.~Geach$^1$, C.~Kobayashi$^1$, S.\,C.\,Chapman$^{2,3}$, C.~Yang$^4$, E.~Gonz{\'a}lez-Alfonso$^5$, J.\,S.~Spilker$^6$, A.~Cooray$^7$, M.\,J.~Michałowski$^8$}

\begin{document}

\maketitle

\begin{affiliations}
 \item Centre for Astrophysics Research, School of Physics, Engineering and Computer Science, University of Hertfordshire, College Lane, Hatfield AL10 9AB, UK
 \item Department of Physics and Atmospheric Science, Dalhousie University, Halifax, NS B3H 4R2, Canada
 \item National Research Council, Herzberg Astronomy and Astrophysics, 5071 West Saanich Rd, Victoria, BC, V9E 2E7, Canada
 \item European Southern Observatory, Alonso de Córdova 3107, Casilla, 19001 Vitacura, Santiago, Chile
 \item Universidad de Alcal{\'a}, Departamento de F{\'i}sica y Matem{\'a}ticas, Campus Universitario, 28871 Alcala{\'a} de Henares, Madrid, Spain
 \item Department of Astronomy, University of Texas at Austin, 2515 Speedway, Stop C1400, Austin, TX 78712, USA
 \item Department of Physics and Astronomy, University of California, Irvine, CA 92697, USA
 \item Astronomical Observatory Institute, Faculty of Physics, Adam Mickiewicz University, ul. Słoneczna 36, 60-286 Pozna{\'n}, Poland
\end{affiliations}

\begin{abstract}

Fluorine is one of the most interesting elements for nuclear and stellar astrophysics\cite{jorissen92, Kobayashi2011}. Fluorine abundance was first measured for stars other than the Sun in 1992$^[$\cite{jorissen92}$^]$, then for a handful metal-poor stars\cite{cunha03}, which are likely to have formed in the early Universe. The main production sites of fluorine are under debate and include asymptotic giant branch (AGB) stars, $\nu$-process in core-collapse supernovae, and Wolf-Rayet (WR) stars\cite{Forestini1992, Woosley1988, Meynet2000, Renda2004, Kobayashi2011b, Spitoni2018, Grisoni2020}. Due to the difference in the mass and lifetime of progenitor stars, high redshift observations of fluorine can help constrain the mechanism of fluorine production in massive galaxies. Here, we report the detection of HF (S/N = 8) in absorption in a gravitationally lensed dusty star-forming galaxy at redshift \textit{z}=4.4 with $N_{\rm HF}$/$N_{\rm{H_2}}$ as high as $\sim2\times10^{-9}$, indicating a very quick ramp-up of the chemical enrichment in this high-\textit{z} galaxy. At \textit{z}=4.4, AGB stars of a few solar masses are very unlikely to dominate the enrichment. Instead, we show that WR stars are required to produce the observed fluorine abundance at this time, with other production mechanisms becoming important at later times. These observations therefore provide an insight into the underlying processes driving the `ramp-up' phase of chemical enrichment alongside rapid stellar mass assembly in a young massive galaxy. 
\end{abstract}

\begin{figure}
   \centering
   \includegraphics[width=0.5\hsize]{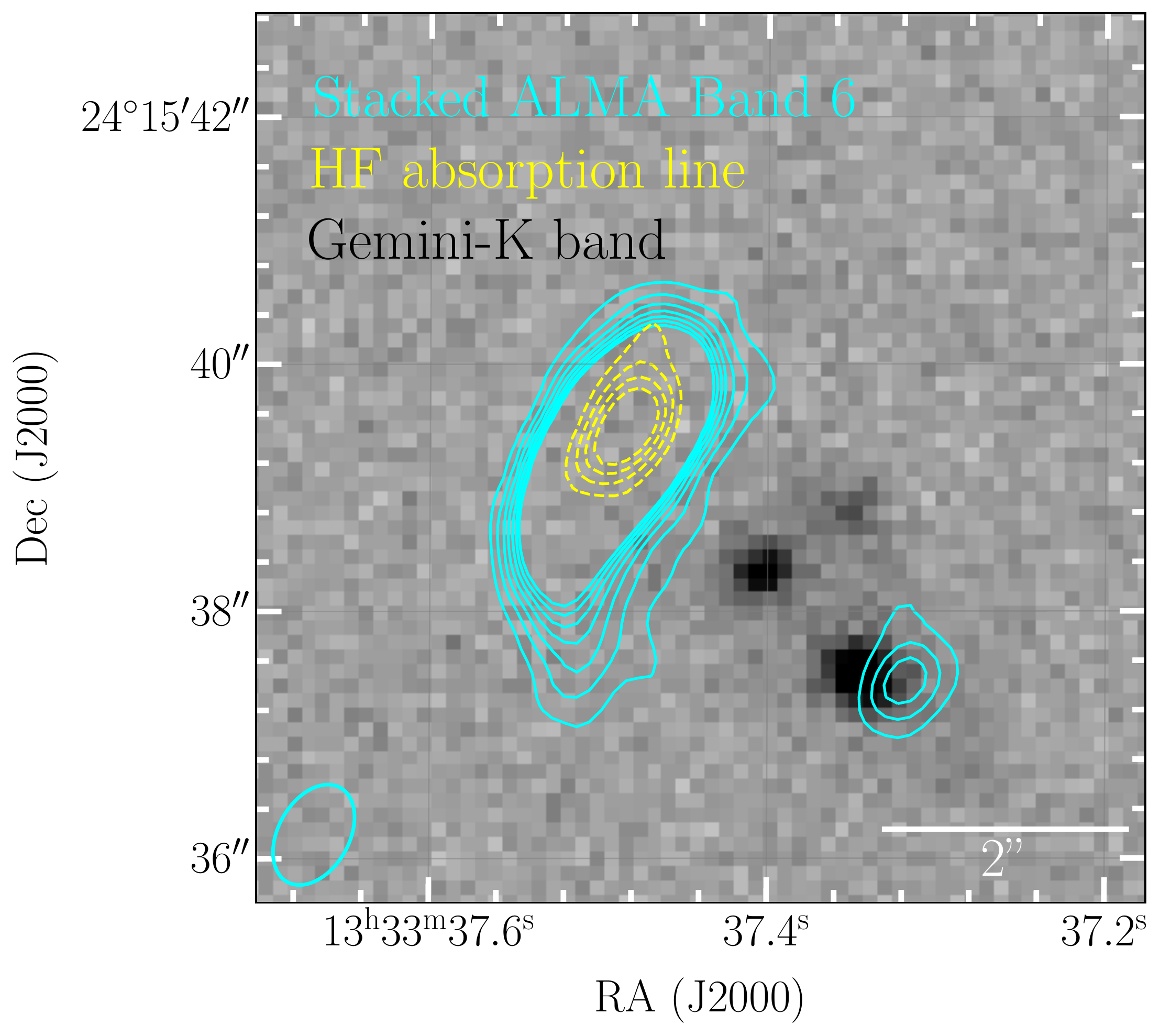}
\caption{\textbf{ALMA contours of NGP--190387}. The contours (4 to 16$\sigma$ in steps of 2$\sigma$) show the 1.36-mm continuum (cyan) and HF absorption line (yellow). The background image is a {\it K}-band image from Gemini/NIRI. The ALMA synthesised beam is shown in the lower left-hand corner.}\label{Fig::Figure1}
   \end{figure}

We observed the gravitationally lensed galaxy NGP--190387 at \textit{z}=4.420 at 211.6--230.0\,GHz (see Methods) with the Atacama Large Millimetre/submillimetre Array (ALMA) seeking to detect the millimetre continuum and various molecular and atomic emission and absorption features in this spectral range. Figure~\ref{Fig::Figure1} shows 1.36-mm continuum contours (r.m.s = 15\,$\mu$Jy\,beam$^{-1}$) overlaid on a {\it K}-band image. The arc-like morphology of the millimetre emission and presence of a counter image with identical emission lines confirm without ambiguity that NGP--190387 is lensed by a foreground structure. Lensing was first suggested based on the spatial offset between the millimetre-wave continuum emission and the sources detected in near-infrared imaging\cite{Fudamoto2017}. The  Gemini Near InfraRed Imager and spectrograph (NIRI) imaging, and data taken in the {\it R} and {\it I}-bands with the Gemini Multi-Object Spectrograph-North (GMOS-N), reveal a group of at least three massive galaxies, most likely members of the foreground lensing structure. The redshift of the lensing group can be estimated from the {\it K}-band Hubble relation\cite{Rocca_Volmerange2004}, suggesting they lie at $z\sim1$--$1.5$.

\begin{figure}
   \centering
      \includegraphics[width=1\hsize]{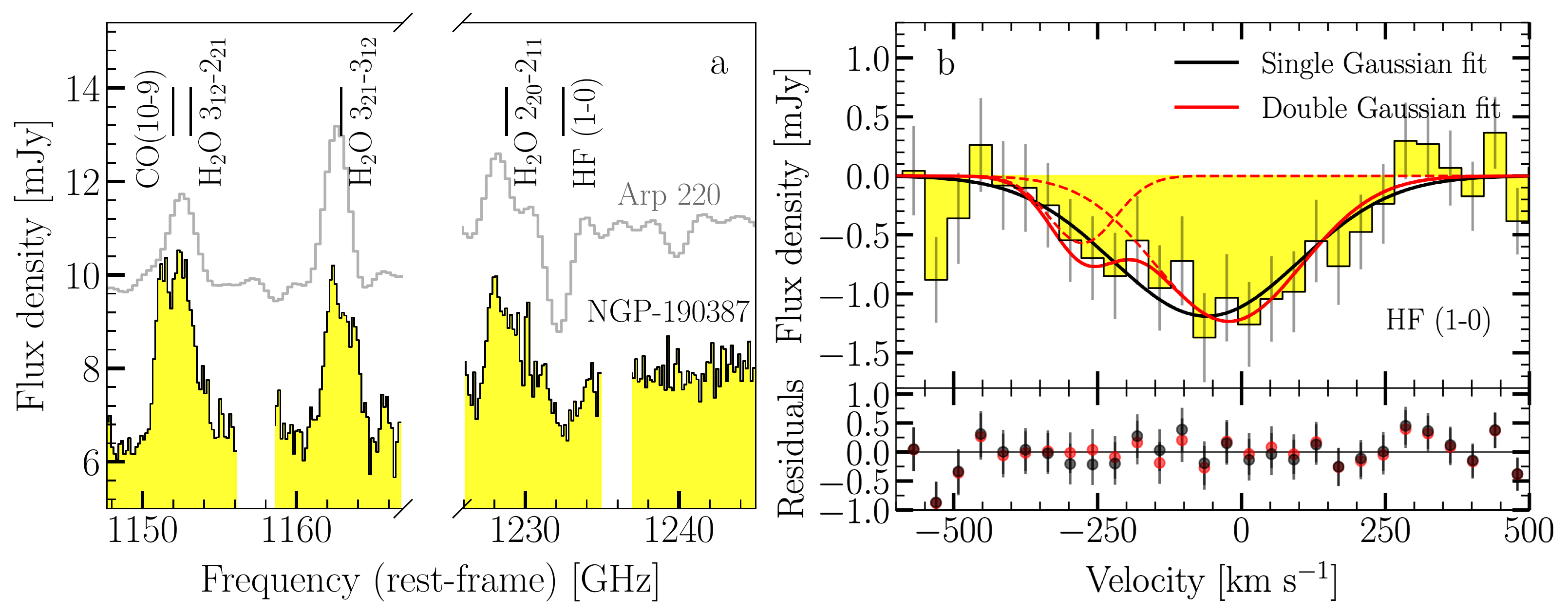}
\caption{\textbf{ALMA spectrum of NGP--190387}. \textbf{a:} Spectrum in Band 6 without continuum subtraction. The position of emission and absorption lines are indicated by vertical lines and  labelled. For comparison, the Arp~220 spectrum\cite{Rangwala2011}, shifted and scaled by arbitrary units is also displayed in gray. The frequency gap is due to the separation of the ALMA  spectral windows. By inspection, the plot of the flux density without continuum subtraction shows that the absorption line is not saturated. By integrating the S/N over the absorption line, we obtain an S/N = 8. \textbf{b:} Observed HF(1-0) line profile ($\nu_{\rm rest}=1232.48$\,GHz$^[$\cite{Pickett1998}$^]$), continuum subtracted. The spectral resolution is 40\,km\,s$^{-1}$, with the velocity scale centered on the HF(1--0) transition at $z=4.420$. The single and double component Gaussian fits are displayed in black and red respectively, with dashed lines indicating the individual components of the double Gaussian fit. The error bars show the uncertainties on the flux measurement given by \texttt{imfit}. The bottom panel shows the residuals (data - model) in units of mJy in black for the single Gaussian fit and in red for the double Gaussian fit. In theses two panels, the flux density is not corrected by the magnification factor ($\mu \approx 5$).}\label{Fig::Figure2}
   \end{figure}

We detect a strong absorption profile corresponding unambiguously to the $J=1\rightarrow0$ transition of HF at $\nu_{\rm rest}=1232.48$\,GHz, redshifted to $\nu_{\rm obs}=227.39$\,GHz at $z=4.420$ (Figure~\ref{Fig::Figure2}). Other emission features are also present, mostly from H$_2$O. We show the line-averaged absorption contours in Figure~\ref{Fig::Figure1}. In Figure~\ref{Fig::Figure2}, we present the continuum-subtracted millimetre-wave spectrum centred at HF(1--0). The continuum flux density at the frequency of the line is $\mu S_\nu=7.2\pm0.3$\,mJy, where $\mu$ is the magnification factor, which we estimate to be $\mu\approx5$ (see Methods). The HF(1--0) line can be described adequately by a single Gaussian profile with $\text{FWHM}=416\pm39$\,km\,s$^{-1}$. We measure an equivalent width of ${\rm EW}= 63\pm10$\,km\,s$^{-1}$ for the absorption line and a velocity-integrated line flux of $\mu S\Delta V=-490\pm80$\,mJy\,km\,s$^{-1}$ (uncertainties being computed by Monte-Carlo simulations). The peak of the Gaussian is slightly offset from the expected position of the transition for the systemic redshift (as defined by the emission lines) by $-64\pm21$\,km\,s$^{-1}$. Therefore, the absorption line shows signs of asymmetry (see Figure~\ref{Fig::Figure2}). If we fit the line with a double Gaussian, the centre of the strongest component is close to zero velocity, $-23\pm30$\,km\,s$^{-1}$, and a weaker component is blue-shifted by $-277\pm35$\,km\,s$^{-1}$. HF(1--0) is a potential tracer of molecular outflows\cite{Monje2014,Lehnert2020}, and therefore we could be seeing tentative evidence of a molecular outflow, as has been detected in a high fraction of similarly luminous objects at high redshift\cite{Spilker2018}. However, the reduced $\chi^2$ of the double Gaussian fit is not significantly better than the single Gaussian fit and could be affected by differential lensing. For these reasons, we use the simple single Gaussian fit in our analysis.

\begin{figure}
   \centering
   \includegraphics[width=0.5\hsize]{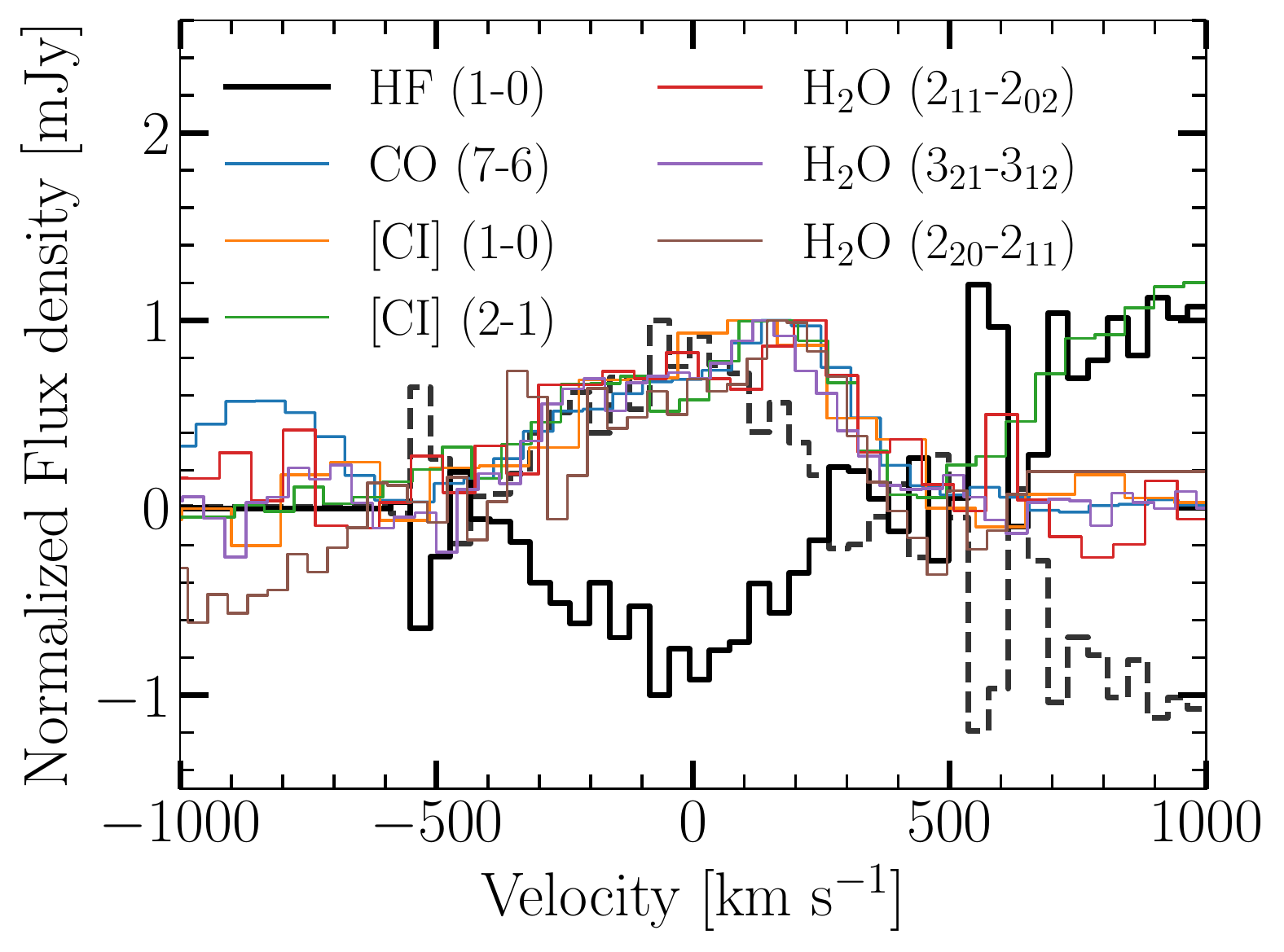}
\caption{\textbf{Overlay of the different ALMA lines}. We displayed the different lines present in the three ALMA bands, omitting CO(10--9) and H$_2$O(3$_{12}$-2$_{21}$) due to blending and H$_2$O$^+$ lines because of the low signal-to-noise. The pattern observed for the HF line at a velocity $>$ 500\,km\,s$^{-1}$ corresponds to the line H$_2$O(2$_{20}$-2$_{11}$) (see Figure~\ref{Fig::Figure2}, left panel). For comparison, we also display the HF(1--0) absorption line, normalised and multiplied by (-1) with a dashed black line.}\label{Fig::Figure3}
   \end{figure}

In Figure~\ref{Fig::Figure3}, we velocity-register and normalise other emission lines detected during the same observation in ALMA Bands 3, 4 and 6. These include several water transitions, CO(7--6) and the two atomic carbon transitions \ci\ and \cii. The emission lines all have a very similar shape reminiscent of the double horned profile of a rotating disk, although merger dynamics could also produce this profile. In contrast, the HF(1--0) absorption line has a profile best described by a single Gaussian with a similar full width at zero intensity as the emission lines. Given the similarity of the profiles of the emission line molecular gas tracers, which span a wide range of excitation conditions (e.g.\,\ci{}\ traces diffuse molecular gas while the high-{\it J} CO and water lines trace denser and warmer material), at first it might seem surprising that HF(1--0) has a different line profile. A simple explanation is that in order to be seen in absorption, the HF-laden gas must be in the foreground of an illuminating continuum source. If, as the emission-line profiles hint, the galaxy is a rotating disk inclined to the line-of-sight, then we will preferentially see absorption due to gas at the near-side of the structure, and thus the full kinematic imprint of the disk rotation will not be observed in the HF(1--0) absorption line. 

\begin{figure}
   \centering
\includegraphics[width=1\hsize]{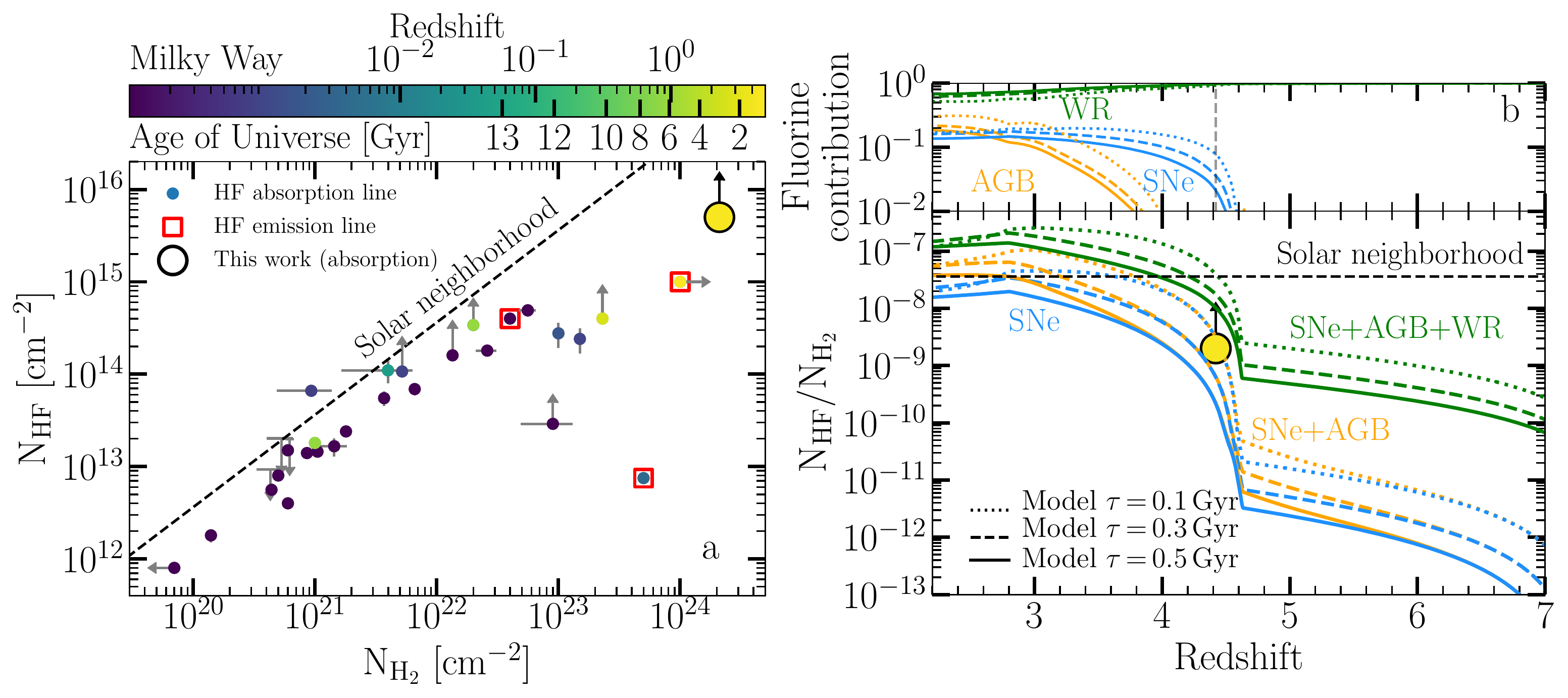}
\caption{\textbf{Evolution of fluorine abundance.} \textbf{a:} HF column density plotted versus H$_2$ column density. We compare our observations to literature data \cite{Neufeld2010, Sonnentrucker2010, Phillips2010, van_der_Werf2010,Rangwala2011, Agundez2011, Monje2011, Emprechtinger2012, Van_der_Tak2012, Kamenetzky2012,Pereira-Santaella2013,Indriolo2013,Monje2014,Kawaguchi2016, Van_der_Wiel2016, Lu2017, Perez-Beaupuits2018, Kavak2019, Lehnert2020} of HF in absorption and emission (red squares) color-coded as a function of redshift. Upper/Lower limits (3$\sigma$) not associated with a data point imply that the HF line has not been detected, while a limit associated with a data point indicates that the method used by the authors may over/underestimate the abundance of the chemical element. \textbf{b:} Top-panel: Contribution of SNe (blue), AGB (orange) and WR stars (green) to the total amount of fluorine (SNe + AGB + WR) predicted by the different chemical evolution models: timescale ($\tau$) = 0.1\,Gyr (dotted line), 0.3\,Gyr (dashed line), 0.5\,Gyr (solid line).
The redshift of NGP--190387 is indicated by a vertical dashed gray line. Bottom-panel: Fluorine abundance of NGP--190387 put into perspective with different chemical evolution models. For the different models the blue, orange, and green lines take into account the production of fluorine by the SNe only, SNe and AGB stars, and SNe, AGB and WR stars, respectively. 
With a timescale longer than 0.1\,Gyr, it would be necessary to have WR stars to reproduce the measured fluorine abundance.
This shows that Fluorine is already present in the interstellar medium at very high redshift and WR stars could be the main Fluorine producer during the key formation epoch of massive galaxies in the early Universe. In both panels, the Solar neighbourhood number ratio of 3.63 $\times$10$^{-8}$ $^[$\cite{Asplund2009}$^]$ is indicated with a dashed line and NGP--190387 is displayed as a large circle outlined in black.}
\label{Fig::Figure4}
   \end{figure}

Aside from the interest in this element from a chemical evolution perspective, HF is potentially an excellent tracer of molecular hydrogen in the diffuse interstellar medium (ISM)\cite{Neufeld2005, Sonnentrucker2010, Emprechtinger2012, Kavak2019} and could be effective to calibrate other tracers over a large range of conditions. Indeed, the main channel for HF formation is through a direct exothermic reaction of fluorine with molecular hydrogen: \smash{\ce{H$_2$ + F  -> HF + H}},  with fluorine being the only element that reacts with H$_2$ in this way\cite{Gerin2016}.  Once formed, HF is difficult to destroy under most interstellar conditions even in low-density environments\cite{Neufeld2005} and is therefore the dominant gas-phase reservoir of fluorine\cite{Neufeld2005}, representing nearly 100\% of gas-phase fluorine over a wide range of temperatures and densities\cite{Neufeld1997}. Despite the relatively low abundance of fluorine, the extraordinary stability of HF allows the accumulation of large amounts of this molecule in the ISM. Since the first detection of HF(2--1) in absorption towards Sagittarius B2\cite{Neufeld1997} with the {\it Infrared Space Observatory}, HF(1--0) has been observed several times in absorption and in emission in the Milky Way. HF has also been detected mostly in absorption in a handful of other local galaxies (see Figure~\ref{Fig::Figure4}-left panel). At higher redshifts, observations are scarce: the molecule has only been detected quasars; in absorption in PKS\,1830$-$211$^[$\cite{Kawaguchi2016}$^]$ at $z=0.9$,  in the Cloverleaf\cite{Monje2011} at $z=2.6$ and in emission in BR\,1202$-$0725 at $z=4.7^[$\cite{Lehnert2020}$^]$. This makes NGP--190387, in addition to being the most distant detection observed in absorption, the only detection outside of the local Universe in a non-quasar host.

We observe both HF absorption and \ci{} emission in this galaxy. While we must take care to consider the radiative transfer effects alluded to above, we can use this information to estimate the HF/H$_2$ abundance, since \ci{} offers an independent tracer of H$_2$$^[$\cite{Papadopoulos2004b}$^]$. We derive $N_{\rm HF} > 5 \times 10^{15}$\,cm$^{-2}$ and  N$_{\mathrm{H_2}}$ = (2.1 $\pm$ 0.4) $\times 10^{24}$\,cm$^{-2}$ (see Methods). This implies an abundance ratio $N_{\rm HF}/$ $N_{\rm H_2} >$ 2 $\times 10^{-9}$.

We compare our measurement with others from the literature, spanning Galactic studies to the distant Universe, in Figure~\ref{Fig::Figure4}-left panel. With the caveat that our HF abundance should be taken as a lower limit, the ratio $N_{\rm HF}/N_{\rm H_2}$ in NGP--190387 may be up to one order of magnitude below that measured for the Solar neighbourhood\cite{Asplund2009} and in diffuse clouds within the Milky Way\cite{Sonnentrucker2010, Indriolo2013, Pereira-Santaella2013}. It is important to consider what internal processes could affect the observed abundance ratio. While at low redshift and low H$_2$ column density ($N_{\rm{H}_{2}}$ $<$ 10$^{23}$\,cm$^{-2}$), the ratio $N_{\rm{H}_{2}}/N_{\rm{HF}}$ seems to follow a clear trend with a low scatter (consistent with the trend found in the ``diffuse cloud abundance'',  1.8 $\times$ 10$^{-8}$ $^[$\cite{Neufeld2005}$^]$), at high redshift and high H$_2$ column density, the trend is less clear. Although it is difficult to draw conclusions with so few robust detections of HF at high redshift, it is interesting to note a possible plateau in the HF column density at high ${\rm H_2}$ density, with $N_{\rm HF} \sim 5\times 10^{14}$\,cm$^{-2}$. NGP--190387 could be the first high redshift example for which an abundance consistent with that found in the diffuse clouds of the local Universe cannot be excluded. Investigations of a larger number of high redshift galaxies at several periods of their evolution will be needed to understand the abundance of fluorine at high H$_2$ column density. It has been suggested that the $N_{\rm HF}/N_{\rm H_2}$ ratio can decrease sharply in dense environments\cite{Emprechtinger2012}, possibly due to accretion (including freeze-out) of HF onto dust grains\cite{Neufeld2005}. This phenomenon might be prevalent in the high density, dusty environment of the ISM in galaxies such as NGP--190387$^[$\cite{Neufeld2010}$^]$. If the low ratio of $N_{\rm HF}/N_{\rm H_2}$ is only due to depletion, that is, the observed relative abundance of the two components is only due to evolution of the gas-phase, we can estimate that the fluorine nuclei in the gas phase to HF molecules in the solid phase ratio, $R = n_{\rm F,gas}/n_{\rm HF, solid}$\cite{Neufeld2005}, is less than 20\% in NGP--190387. In this scenario, the main fraction of fluorine in NGP--190387 would be locked up in dust. However we think this level of freeze-out unlikely as the Cosmic microwave background (CMB) temperature at the redshift of the source is $T_{\rm CMB}\approx$ 15\,K, and dust is likely to be in the vicinity of an intense starburst (see Methods).

A simpler explanation for the relatively low HF abundance relative to H$_2$ compared to Solar is the picture where early galaxies are gas-rich, with high molecular gas fractions\cite{Tacconi2018} and star-formation rates, but enrichment of the ISM is still ramping up. The origin of fluorine is still poorly understood and not well constrained, and so-far, fluorine nucleosynthesis processes have only been observationally confirmed in AGB stars\cite{Kobayashi2011}. The formation of fluorine in AGB stars during the thermally pulsing phase depends on the stars' initial masses and metallicity, with a production that is higher for initial masses 2--4\,M$_\odot$$^[$\cite{Lugaro2004}$^]$ with a peak at $\sim$3\,M$_\odot$$^[$\cite{Kobayashi2011}$^]$. AGB stars can produce more than 50\% of the total fluorine in the Solar neighbourhood\cite{Kobayashi2020}. The lifetimes of 2--4\,M$_\odot$ stars are about 0.2--0.9\,Gyr at low metallicities. It is expected that the F/H ratio is much lower than the Solar, since F is produced from low-mass stars during their AGB phase. If the star formation history of NGP--190387 were similar to our Milky Way Galaxy, the fluorine abundance would be $\sim 1\%$ of the Solar neighbourhood abundance at redshift of 4.4, since there is not enough time to have the enrichment not only from AGB stars but also from massive stars. However, this galaxy experience an episode of intense star formation (see Methods). With a timescale of less than 0.1 Gyr, it could be possible to reproduce the observed F/H with a rapid enrichment from massive stars via supernovae. If the star formation timescale is longer, then it would be necessary to have extra sources, which are WR stars, i.e., rotating massive stars.

At $z=4.4$ the contribution to fluorine production by AGB stars is therefore small, unless \textit{(i)} the formation redshift is very high ($z>10$; when the Universe was 450~Myr old, 900 Myr before \textit{z}=4.4), \textit{(ii)} a significant fraction of stars have formed almost instantaneously, and \textit{(iii)} this galaxy is already chemically evolved with the stellar metallicity greater than $0.1\,Z_\odot$ (see Methods). However, this problem can be solved with the inclusion of WR stars. In Figure~\ref{Fig::Figure4}-right panel, we show the predictions of state-of-the-art chemical evolution models with and without WR contribution. The evolution of the fluorine abundance, $\rm{[F/H]}\sim \log_{10} [N_{\rm HF}/N_{\rm H_2}/(N_{{\rm F},\odot}/N_{{\rm H},\odot})]$, in the ISM of NGP--190387 is calculated with so-called one-zone model\cite{Kobayashi2000}, which includes the metal-dependent nucleosynthesis yields of the full mass range of stars.

In the standard model with nucleosynthesis yields that are consistent with the chemical evolution of the Milky Way, the first and second peaks of the fluorine abundance evolution (that we can see at $z\sim4.4$ and $z\sim3$ in the Figure~\ref{Fig::Figure4}-right panel) are caused by core-collapse supernovae (13--50\,M$_\odot$$^[$\cite{Kobayashi2020}$^]$) and AGB stars (1--8\,M$_\odot$$^[$\cite{Kobayashi2011}$^]$), respectively (we note that the contribution from super-AGB stars (8--10M\,$_\odot$) is also included in the models but they make no difference\cite{Kobayashi2020}). With only AGB stars, the model fluorine abundance is too low to explain the observations.
In addition to the standard model, we also show a model including stellar rotation, which predicts a significant contribution from WR stars even at very low metallicities. In this case, pre-supernova and core-collapse supernova yields are taken from the literature\cite{limongi18} (set ``R'' , for set Recommended\cite{limongi18}) with stellar rotation of 300\,km\,s$^{-1}$ at 13--120\,M$_\odot$, in addition to the same yields for super-AGB and AGB yields as in the standard model, and considering that all stars with mass greater than 25\,M$_\odot$ fully collapse to a black hole (except for pre-supernova mass-loss).
This contribution from WR stars should be considered as a maximum as the high rotation velocity is assumed for all massive stars regardless of metallicity; note that in  observations of nearby stars, the peak of rotational velocity is much smaller\cite{ramirez13}. It is also important to note that the fluorine abundance also depends on the stellar initial mass function (IMF), and a Kroupa IMF with a slope of $x=1.3$ is adopted for 0.01--120\,M$_\odot$ in these models. A more bottom-heavy IMF would give enhanced AGB contribution but a lower contribution from supernovae and WR stars, resulting in lower F abundances than in these models. The flattest IMF with the slope of $x=1.1$ gives a $\sim$0.4\,dex higher abundance, which is also not large enough to explain the observation unless the star formation timescale is extremely short.

The predicted fluorine abundance with the inclusion of WR stars increases very quickly to reach the observed value of NGP--190387, and stays high afterwards. This may add further observational evidence (and the first outside the local Universe) of the need of WR stars in the nucleosynthesis of fluorine and further strengthens the hypothesis that WR stars (and/or $\nu$-process\cite{Kobayashi2011b}) are a dominant channel of ISM chemical enrichment during the key formation epoch of massive galaxies in the early Universe. Novae can also be fluorine producers\cite{Wiescher1986} but do not contribute in our high-redshift galaxy due to the long timescale.

\begin{methods}

\subsection{Galaxy selection}
HATLAS\,J133337.6+241541, hereafter NGP--190387, was discovered\cite{Ivison2016} in the H-ATLAS survey\cite{Eales2010} (\textit{Herschel} Astrophysical Terahertz Large Area Survey), which was undertaken with the Spectral and Photometric Imaging Receiver\cite{Griffin2010} (SPIRE) on-board the {\it Herschel Space Observatory}\cite{Pilbratt2010}. NGP--190387 was identified as a `500-$\mu$m riser'\cite{Cox2011}, with $S_{250}$\,$<$\,$S_{350}$\,$<$\,$S_{500}$, which identifies potential high-redshift ($z>4$) dusty star-forming galaxies (DSFGs). The redshift ($z=4.420\pm0.001$) and astrometric position (13$^{\rm h}$33$^{\rm m}$37.47$^{\rm s}$, +24$^{\circ}$15$'$39.3$''$ J2000) of NGP--190387 were measured unambiguously using the NOrthern Extended Millimetre Array (NOEMA) through $^{12}$CO(5--4) and (4--3) line emission\cite{Fudamoto2017}.  This study detects multiple additional emission lines at the same position and redshift.

\subsection{Data reduction}~The ALMA observations were carried out on 20 March 2018 during the ALMA Cycle 5 campaign (project 2017.1.00510.S). The observations used 46 antennas of the 12\,m array, in configuration C43--4 in Band 6. The target was observed for 61.5 min (on-source integration time 42.3 min) to reach a median r.m.s.\ per 40-km\,s$^{-1}$ channel of 0.14\,mJy\,beam$^{-1}$ in Band 6. The full width half maximum (FWHM) synthesised beam size is 0\farcs88 $\times$ 0\farcs57 at  position angle 29.5\degree. J1229+0203 and J1327+2210 were observed as calibrators; the flux density scale should be accurate to 5 per cent. The data were calibrated using the Common Astronomy Software Applications (CASA) ALMA pipeline v.5.6.1-8$^[$\cite{McMullin2007}$^]$. We used CASA task \texttt{tclean}, employing multi-scale cleaning using a geometric progression of scales\cite{Cornwell2008}, with natural weighting to maximise sensitivity, masking the central region. Continuum subtraction was performed in UV-space using the line-free channels with the CASA task \texttt{uvcontsub}. New continuum subtracted cubes have been created using the same procedure as before. Galaxy sizes (including error estimations\cite{Condon1997}) were obtained using the CASA task \texttt{imfit} in each channel. The sizes presented in this paper are the deconvolved sizes, uncorrected for magnification factor unless otherwise stated. For a S/N of $\sim$3 per channel, the reliable size measurement limit for an interferometre is approximately half the beam size\cite{Marti-Vidal2012}.

\subsection{Infrared properties}~We estimate a rest-frame 8--1000-$\mu$m infrared luminosity, uncorrected for gravitational amplification, of ${\log_{10}}(\mu L_{\rm IR}/L_\odot)=13.5\pm0.2$ from the SED-fitting code \texttt{CIGALE}\cite{Boquien2019}. We measure the dust temperature by fitting a modified blackbody model with emissivity $\beta=1.5$$^[$\cite{Kovacs2006,Gordon2010}$^]$, $T_{\rm dust}=42\pm 2$\,K. Even for the highest plausible magnifications, this suggests a prodigious star-formation rate, $\text{SFR} \approx 3500/\mu$\,M$_\odot$\,yr$^{-1}$, for a Kroupa initial mass function\cite{Kennicutt1998,Kroupa2001}.

\subsection{Details on the computation of N$_{\rm{H}_{2}}$}

To determine $N_{\rm H_2}$ we use the \ci\ emission line and assumed a C\,{\sc i}/H$_2$ conversion factor $X_{\mathrm{C\,{\sc I}}}$:
\begin{equation}
N_{\mathrm{H}_{2}}~[\mathrm{cm^{-2}}] = X_{\mathrm{CI}}~\mathrm{[cm^{-2}~(K~km\,s^{-1})^{-1}]} \times W_{\mathrm{CI}}~[\mathrm{K~km\,s^{-1}}]
\end{equation}
\noindent where $N_{\mathrm{H}_{2}}$ is the $\mathrm{H}_{2}$ column density and $W_{\mathrm{CI}}$ is the integrated brightness temperature of the \ci\ transition. Integrating over the source we measure $W_{\mathrm{CI}} = (6.4\pm1.2)\times 10^{3}$\,K\,km\,s$^{-1}$ (the uncertainties were calculated performing 10\,000 Monte-Carlo simulations
perturbing the measurement randomly within the uncertainties). We assume $X_{\mathrm{CI}}=$ $(3.3\pm0.2)\times 10^{20}$~cm$^{-2}$\,(K\,km\,s$^{-1}$)$^{-1}$ from the ratio between the H$_2$ mass and the line luminosity L$^\prime_{\mathrm{C\,{\sc I}(1-0)}}$:
\begin{equation}
X_{\mathrm{CI}}~\mathrm{[cm^{-2}~(K~km\,s^{-1})^{-1}]} = \frac{\rm M_{\rm{H2}}~\mathrm{[M_\odot]}}{\rm L^\prime_{\mathrm{C\,{\sc I}(1-0)}}~\mathrm{[K\ km\ s^{-1}\ pc^2]}} \frac{1}{1.6 \times 10^{-20} ~\mathrm{[M_\odot\ pc^{-2}\ cm^{2}]}}
\end{equation}
\noindent with $\mu$L$^\prime_{\mathrm{C\,{\sc I}(1-0)}} = (6.0 \pm 0.3) \times 10^{10}$ K\ km\ s$^{-1}$\ pc$^2$, $\mu$M$_{\rm{H2}} = (3.2 \pm 0.2) \times 10^{11}$ M$_\odot$, using\cite{Papadopoulos2004a, Wagg2006}:
\begin{equation}\label{Eq::H2fromCI}
 M_{\rm H_2}\ [M_{\odot}]= 1375.8 \frac{d_L^2}{1+z} \left(\frac{X^{\prime}_{\mathrm{C\,{\sc I}(1-0)}}}{10^{-5}}\right)^{-1}  \left(\frac{A_{\mathrm{10}}}{10^{-7}}\right)^{-1} Q_{10}^{-1}\ S_{\mathrm{C\,{\sc I}}}\Delta v\ [\mathrm{Jy\ km\ s ^{-1}}],
\end{equation}
\noindent with the Einstein coefficient A$_{\mathrm{10}}$  = 7.93$\times$10$^{-8}$s$^{-1}$ , the [C\,{\sc i}]-to-H$_2$ abundance ratio $X^{\prime}_{\mathrm{C\,{\sc I}}}$ = 3$\times10^{-5}$~$^[$\cite{Weiss2003}${^]}$
and Q$_{10}$ = 0.6$^ {[}$\cite{Bothwell2017}${^]}$ the excitation factor. This result has then been divided by 1.36 to remove the contribution from helium. This ratio is consistent with that found for local ULIRGs\cite{Jiao2017} giving N$_{\mathrm{H_2}}$ = $(2.1\pm0.4) \times 10^{24}$\,cm$^{-2}$. Although for nearby galaxies, atomic hydrogen dominates the total mass of gas in galaxies, this proportion is reversed for more distant galaxies due to a strong increase in molecular hydrogen\cite{Daddi2010, Lagos2015,Tacconi2018} and we adopt  $M_\mathrm{H_2} + M_\mathrm{H} \sim M_\mathrm{H_2}$, valid at $z > 0.4$$^[$\cite{Tacconi2018}$^]$.

\noindent
The \ci\ surface brightness of the galaxy ($W_{\mathrm{CI}}$) is given by\cite{Solomon2005}:
\begin{equation}
    W_{\mathrm{CI}}\ [\mathrm{K~km\,s^{-1}}] = \frac{(1+z)\ c^2}{2 k\ \nu_{\rm obs}^2\ \Omega_s} S_{\mathrm{C\,{\sc I}}} \Delta v = 1.38 \times 10^6 \frac{ (1+z)\  S_{\mathrm{C\,{\sc I}}} \Delta v\ [\mathrm{Jy\ km\ s ^{-1}}]}{\nu_{\rm obs}^2  [\mathrm{GHz}]\ \Omega_s\  [\mathrm{arcsec^2}]}
\end{equation}

\noindent where $\nu$ is the observed frequency, $k$ the Boltzmann constant, $\Omega_{s}$ the source solid angle ($\Omega_{s}$\,=\,$\pi\theta_{\rm min}\theta_{\rm max}$/ 4$\ln$(2)), and $S_{\mathrm{C\,{\sc I}}}$ $\Delta v$ the velocity integrated flux density. We corrected the observed \ci\ flux for the CMB\cite{daCunha2013} by assuming local thermodynamic equilibrium (LTE) between the gas and the dust. 
This represents 24$\pm$2\% of the flux that had to be corrected (added) in the column density calculation above. The size of the \ci\ emission line is too uncertain in Band 3 to be measured accurately, the beam size in Band 3 is 1.39'' $\times$ 0.95''. We took advantage of the higher resolution in Band 6, to estimate the size of the emission line. We assume that the CO(10-9) emission line traces the same material as \ci{}$^[$\cite{Ojha2001, Ikeda2002,Papadopoulos2004b}$^]$ (both lines show similar profiles) and its size can therefore be used to calculate the H$_2$ column density. In addition, because HF has the same spatial distribution as CO(10-9) (see Details on the computation of N$_{\rm HF}$), they are both reasonable tracers of the spatial distribution of the gas responsible for the emission/absorption. The CO(10-9) emission line being partially blended with the emission line of water H$_2$O(3$_{12}$-2$_{21}$) (see Figure~\ref{Fig::Figure2}-left panel), and since there is no significant variation in size between these two components, we have considered the average size of these blended elements. We measure a source size of $(0.70''\pm0.05'') \times (0.28''\pm 0.04'')$, the uncertainties corresponding to the standard deviation of the sample means. For ULIRGs, there may be a fine structure deficit. For this reason, we also calculated N$_{\rm H_2}$ from the continuum and water emission lines of NGP--190387. The results from these different techniques give comparable results.

\subsection{Details on the computation of N$_{\rm HF}$}

For extragalactic sources the HF absorption is generally expected to be physically associated with the source(s) of far-infrared continuum, unless the galaxy is seen edge-on or there is a foreground component in the direction of the continuum source. We note that in the case of NGP--190387, the \ci\ profile that is used to estimate N$_{\rm H_2}$ mimics the profiles of the excited CO and H$_2$O lines (Figure~\ref{Fig::Figure3}), so that there is no evidence for detached foreground low-excitation gas. The HF(1-0) absorption profile is slightly blueshifted ($\sim$ 60 km s$^{-1}$) relative to the emission [C\,{\sc i}], CO and H$_2$O profiles, but this is a rather common characteristic of local ULIRGs where the excited OH 65$\mu$m absorption, radiatively pumped by the nuclear far-IR field, is also slightly blueshifted relative to the [C\,{\sc ii}] 158$\mu$m emission line\cite{Gonzalez_Alfonso2017}. Furthermore, the velocity coverage of HF(1-0) is the same as that of the CO and H$_2$O lines.

If the HF is physically associated with the continuum, we should then consider the excitation effects of the infrared radiation on HF(1–0), and also the fact that only the HF in front of the continuum will produce absorption. In order to consider these effects and quantify their impact on the estimation of N$_{\rm HF}$, we have modeled the source using the non-LTE, non-local approach\cite{Gonzalez_Alfonso2014}, where dust and molecules are assumed to be mixed. Any geometrical effect is ignored by considering a spherically symmetric source; even if the actual geometry is a disk, it can be considered in a first approach as an ensemble of these spherical clouds, and EW$_{\rm{HF (1-0)}}$ would be the same for the individual clouds and for the ensemble.

By taking the following parameters in the model: a dust temperature of 50 K, $\tau_{100}$ = 2, corresponding to N$_{\rm H_2} \sim 1.3\times$10$^{24}$ cm$^{-2}$ (consistent with the value found by the \ci \ line) for gas-to-dust ratio of 100, V$_{\rm{turb}}$ = 240 km s$^{-1}$ and by excluding the effects of the collisions, we tested two different values of the HF column density: N$_{\rm HF}$ = 5$\times$10$^{14}$ cm$^{-2}$ and N$_{\rm HF}$ = 5$\times$10$^{15}$ cm$^{-2}$. The absorption produced in the N$_{\rm HF}$ = 5$\times$10$^{14}$ cm$^{-2}$ model is very weak, with only EW$_{\rm{HF (1-0)}}$ = 8 km~s$^{-1}$. In contrast, the N$_{\rm HF}$ = 5$\times10^{15}$ cm$^{-2}$ model gives EW$_{\rm{HF (1-0)}}$ = 66 km s$^{-1}$, in agreement with the observations. Since a lower value of V$_{\rm{turb}}$ (i.e. a profile broadened by systemic motions rather than by turbulence alone) would increase the required N$_{\rm HF}$, the HF column density of $5\times10^{15}$ cm$^{-2}$ would be still considered a lower limit; we conservatively estimate an abundance of $2\times10^{-9}$ with 0.4 dex of uncertainty (Figure~\ref{Fig::Figure4}).

We also measured the deconvolved size of the HF(1--0) absorption using the CASA task \texttt{imfit} after subtraction of the continuum. In order to be consistent with the emission line size measurements, we multiplied the flux density by -1. Because of the weaker signal of the HF(1--0) absorption line compared to the emission lines (see Figure~\ref{Fig::Figure2}-left panel), the size measurement could only be made on fewer channels and may suffer from more uncertainty. We measure a source size of $(0.59''\pm0.06'') \times (0.19''\pm 0.07'')$, the uncertainties corresponding to the standard deviation of the individual sizes across different channels. The measured size of the HF(1--0) absorption line is slightly smaller than the emission line region we consider for the gas. However, as the sizes are consistent within uncertainties, we consider that the assumption of considering that the H$_2$ gas probes the same region as HF is justified.

\subsection{Chemical evolution models}

The fluorine evolution depends on the star-formation history (SFH), in addition to the enrichment source and IMF as already discussed.
This galaxy is extremely faint/not detected across the UV-optical-NIR, and therefore we have only limited constraints on the SFH from stellar population modelling\cite{Boquien2019}. 
We therefore investigated and modelled all possible SFHs varying i) star formation efficiency (SFR/M$_{\rm gas}$), ii) duration of star burst, iii) the fraction of additional, older stellar population, iv) the formation epoch of the old stellar population, v) gas accretion timescale, and vi) the IMF. Based on its observed properties (infrared luminosity, gas mass, SF timescale), this galaxy is similar to submillimetre galaxies (SMGs). We therefore found that iii), iv), and v) are not important for chemical evolution of galaxies.
There is no observational indication that the IMF of this galaxy should be different from that in the Milky Way, and thus we present the models with the Kroupa IMF in Figure~\ref{Fig::Figure4}-right panel.
We then find that the observed F/H ratio of NGP--190387 can be reproduced (A) if the star formation timescale is shorter than 0.3 Gyr, or (B) with significant F production from WR stars.

In Figure~\ref{Fig::Figure4}-right panel, we present the models with star formation timescales from 0.1 to 0.5 Gyr, which is a relative constant between SFR and gas fraction at a give time (also called depletion timescale)\cite{Kobayashi2000}.
The gas fraction increases due to the accretion of pristine gas with an exponential timescale of 1 Gyr in these models, and almost identical results are obtained with a longer timescale (e.g., 5 Gyr); the accretion timescale would have an impact for a much longer star formation timescale such as in our Milky Galaxy.
The high SFR of this galaxy requires some mechanism of a starburst, such as galaxy merger, and in that case, the duration of starburst is expected to be as short as 0.1 Gyr.
We also assume the formation epoch to be the epoch of reionization, $z =$ 7.7$^[$\cite{Planck2020}$^]$, which is most plausible; an earlier formation epoch (e.g., $z=10$) results in a slightly more old stellar population, and hence in a slightly higher F abundance at the onset of the starburst.
However, this contribution from old stellar populations is not important because the F abundance rapidly increases during the starburst as seen in the Figure.
Similarly, increasing the SFR before the starburst (by a factor of 10) does not affect either.
Finally, if the duration of starburst is longer than in the models, the predicted gas fraction becomes too small for this galaxy at this redshift.

Our chemical evolution model includes the production of F from all plausible sources, i.e., core-collapse supernovae, AGB stars (including super-AGB stars), and WR stars using the most recent nucleosynthesis yields\cite{Kobayashi2020}, self-consistently. There is an uncertainty of the F yields due to the nuclear reaction rates and the description of stellar mass loss, rotation, and rotation-induced mixing, which are calibrated by the observations of stars\cite{limongi18}. $^{19}$F is synthesized from $^{14}$N in He convective shell but can be destroyed by the reaction $^{19}$F($\alpha$, $p$)$^{22}$Ne, which is uncertain. The Wolf-Rayet stars might not produce much $^{19}$F $^[$\cite{Palacios2005}$^]$, but noted that their mass-loss rates, and hence F yields, might be underestimated. It is also important to note that a significant production of F from WR stars is required from the Galactic chemical evolution models\cite{Kobayashi2020,Spitoni2018,Prantzos2018} and also the observational of stars\cite{Jonsson2014} in the Milky Way.

\subsection{Determination of the magnification factor}
The determination of the magnification factor has been done in the UV plane using the \texttt{VISILENS}\cite{Hezaveh2013, Spilker2016} package. A detailed description of the lens modeling will be presented in a companion paper. Using continuum emission, having first subtracted the emission lines in Band 6 and taking the hypothesis of a single lens we obtain a magnification factor $\mu=4.9 \pm 0.2$.

\subsection{Freeze-out of the fluorine on dust grains}
One hypothesis to explain the low abundance of hydrogen fluoride in NGP--190387 compared to the abundance in the vicinity of the sun is that HF can condense onto dust grains. We have investigated this possibility by considering that the HF abundance follows from balancing its adsorption onto dust grains and desorption\cite{Van_der_Wiel2016,Rodgers2003,Jorgensen2005}.
 The characteristic time for the HF(1--0) to freeze-out ($\tau_\mathrm{freeze-out}$) is defined by\cite{Van_der_Wiel2016}:
\begin{equation}
\tau_\mathrm{freeze-out}(n_{\mathrm{H}}, T_{\mathrm{gas}})\ [yr] = 6.97 \times 10^{9}\left(\frac{m_{\mathrm{HF}}}{T_{\mathrm{gas}}}\right)^{0.5} n_{\mathrm{H}}^{-1},
\end{equation}
\noindent where $T_{\mathrm{gas}}$ is the temperature of the gas, $m_{\mathrm{HF}}$ is the molecular weight of HF, and $n_{\mathrm{H}}$ is the density of hydrogen nuclei in m$^{-3}$.
The characteristic desorption time ($\tau_\mathrm{desorption}$) is more difficult to estimate because it depends on the type of grain and is expressed as\cite{Van_der_Wiel2016}:
\begin{equation}
\tau_\mathrm{desorption}(T_{\text {dust }})\ [yr] = \frac{3.17 \times 10^{-21}}{\exp \left(-\frac{E_{\mathrm{b}, \mathrm{HF}}}{k T_{\text {dust}}}\right)},
\end{equation}
\noindent where $T_{\mathrm{dust}}$ is the temperature of the dust, $k$ is the Boltzmann constant, and  $E_{\mathrm{b}, \mathrm{HF}}$ is the binding energy of HF to the the dust grain surface. HF(1--0) may condense on to the dust grains if $\tau_\mathrm{freeze-out} <\tau_\mathrm{desorption}$. By taking the most conservative values of the parameters possible: $T_{\mathrm{gas}} \in [35-100K]$, $T_{\mathrm{dust}} \in [35-50K]$ and $n_{\mathrm{H}} \in [10^3-10^5$ cm$^{-3}$], and different values of $E_{\mathrm{b}, \mathrm{HF}}$ corresponding to Hydrogenated crystaline silica, CO ice on amorphous silica, and CO$_2$ ice on amorphous silica\cite{Van_der_Wiel2016}, the freeze-out phenomenon is negligible (the transition is highly dependent on the temperature and occurs at $\sim$ 20K). On the other hand, one possibility could lead to an adsorption of HF on the grains, if they are covered by a pure water layer. This possibility has been studied in the case of NGC 6334 I and I(N) and has been rejected\cite{Van_der_Wiel2016}. While being well aware that the conditions of the interstellar medium of NGP--190387 are not homogeneous and may suffer from large local disparities, our conclusion is that freeze-out does not explain the low HF abundance relative to the Solar abundance.

\subsection{Data Availability}
This paper makes use of the following ALMA data: 
\newline
ADS/JAO.ALMA\#2017.1.00510.S, archived at \href{https://almascience.nrao.edu/alma-data/archive}{https://almascience.nrao.edu/alma-data/archive}. Tabulated spectral data used in this study is provided at 
\newline
\href{https://github.com/maximilienfranco/f21\_fluorine\_spectrum}{https://github.com/maximilienfranco/f21\_fluorine\_spectrum}

\subsection{Code Availability}
The ALMA data are processed using the CASA ALMA pipeline (v.5.6.1-8) available at \href{https://almascience.nrao.edu/processing/science-pipeline }{https://almascience.nrao.edu/processing/science-pipeline}. The lens model was done using the \texttt{VISILENS} package publicly available at \href{https://github.com/jspilker/visilens/}{https://github.com/jspilker/visilens/}

\subsection{Acknowledgements}

\end{methods}

\begin{addendum}
 \item M.F. is grateful to Yashar Hezaveh for his advice on the lens model. M.F. and K.E.K.C. acknowledge support from the UK Science and Technology Facilities Council (STFC) (grant number ST/R000905/1). K.E.K.C. acknowledges support from a Royal Society Leverhulme Trust Senior Research Fellowship (grant number RSLT SRF/R1/191013). J.E.G. acknowledges support from a Royal Society University Research Fellowship. C.K. acknowledges funding from the UK STFC through grant ST/M000958/1 \& ST/ R000905/1. S.C.C. acknowledges the Natural Sciences and Engineering Research Council of Canada (NSERC). C.Y. acknowledges support from an ESO Fellowship. J.S. is a NHFP Hubble Fellow supported by NASA Hubble Fellowship grant no. HF2-51446 awarded by the Space Telescope Science Institute, which is operated by the Association of Universities for Research in Astronomy, for NASA, under contract NAS5-26555. M.J.M. acknowledges the support of  the National Science Centre, Poland through the SONATA BIS grant 2018/30/E/ST9/00208. This paper makes use of the following ALMA data: ADS/JAO.ALMA\#2017.1.00510.S. ALMA is a partnership of ESO (representing its member states), NSF (USA) and NINS (Japan), together with NRC (Canada), MOST and ASIAA (Taiwan), and KASI (Republic of Korea), in cooperation with the Republic of Chile. The Joint ALMA Observatory is operated by ESO, AUI/NRAO and NAOJ.
 \vspace{-0.3cm}
  \item[Author contributions] M.F. reduced and analysed the data; M.F., K.E.K.C., J.E.G., C.K. interpreted the results and wrote the paper. C.K. created the chemical evolution models. C.Y. interpreted the results. J.S. helped create the lens model. E.G.-A. computed the HF column density and contributed to various aspects of the analysis. S.C.C. provided the data. All other authors contributed to the ALMA proposals, to the scientific discussion, and provided comments to the manuscript.
 \item[Competing Interests] The authors declare that they have no competing interests.
  \vspace{-0.3cm}
 \item[Correspondence] Correspondence and requests for materials
should be addressed to M.\,Franco~(email: \href{mailto:m.franco@herts.ac.uk}{m.franco@herts.ac.uk}).
 \vspace{-0.3cm}

\end{addendum}

\newpage

\end{document}